# DISCRETE TUNNELING IN GRANULATED SUBSTANCES AND OTHER SIMILAR MEDIUMS

E. S. Demidov, N. E. Demidova

*Nizhniy Novgorod state university, Nizhny Novgorod, Russia*
demidov@phys.unn.ru

Work is devoted to physics of current transport in a wide class of the hetero-phase granulated mediums and similar systems with set of metal or semi-conductor granules, quantum dots or potential wells in which the exit from Coulomb blockade tunneling regime can be not observable because of irreversible breakdown and destruction of structure of medium. Such systems also concern and the condensed mediums with short distanced atoms of transition elements. In article for small and average electric fields the analytical decision of a stationary problem of discrete electronic transport through a chain of as much as big number of metal granules in area Coulomb tunneling blockade is performed. It is deduced the exponential law of growth of a current with electric field in such granulated systems. The characteristic feature of discrete tunneling in such medium is the volt-ampere characteristic type $I \sim \exp(V/(N+1)kT)$ with great value $N \gg 1$. Examples of application of the theory for explanation of current transport in porous silicon, synthesised by ionic implantation of nitrogen in silicon layers of nitride of silicon or glass like amorphous semiconductors are resulted.

## 1. INTRODUCTION

Usually by consideration of features Coulomb blockade tunneling of electrons through conducting granules or quantum points in the dielectric substances between metal electrodes the basic attention is given to current steps on volt-ampere characteristics (VAC) or to conductivity peaks in such structures depending on quantity of granules, their parameters and a tunnel transparency of barriers [1-4]. Interest to VAC features of systems with a small amount of granules or quantum dots is caused by prospects of development of promising single-electronics [5]. At the same time the physics of electron current transport in a wide class of the heterophase granulated mediums and similar systems with the big set of granules, quantum dots or potential zero-dimensional wells in which the exit from Coulomb blockade tunneling regime can be not observable because of irreversible breakdown and destruction of structure of medium is of interest. Irreversibility can be caused by instability of uniform distribution of density of current, current cording because of characteristic for discrete tunneling of its superlinear growth with electric field and disorder of local parameters of medium. These mediums concern nanodispersed metal-dielectric composite materials, intensively investigated recently semiconductor epitxial heterostructures with set of quantum dots and porous silicon. Such structures can be glass like amorphous semiconductors and also, according [6], the containing close located atoms of transition metals dielectrics. In work [6] the attention to possibility of discrete tunneling электронов through atoms of transition metals with extremely small electric capacitance has been paid

$$C = (\partial^2 E / \partial Q^2)^{-1} = e^2/U,$$

where $Q=en$, $e$ - electron charge, $E$ - energy of atom with not completely filled $d^n$- or $f^n$-shell containing $n$ electrons, $U$ - energy interval between two neighbor recharge levels of atom.

It is obvious, that the problem of discrete tunneling in the three-dimensional medium is extremely difficult. In the problem of Coulomb blockade tunneling in system with big number of tunnel junctions only the one-dimensional problem solving for discrete tunneling through a chain of as much as big number of the identical metal electrodes divided by identical tunnel transitions is known [7]. The electrostatic contribution was modeled by the equivalent scheme from series connection of electric capacitances C of tunnel junctions and parallel capacitances $C_0$ between each electrode and the certain common ground electrode. With application of the numerical Monte-Carlo method it is shown, that the electric voltage enclosed by the ends of a chain, raise in it one-electronic charging solitons which can transfer in the chain and interact with each other. However, as in case of small number of granules, in article [7] the basic attention was given to the big excitations of system by external electric field. Presented in [7] results of numerical simulating do not allow to use them in the case of extremely small tunnel currents interesting us. Results of the solving for small currents have not been resulted. Besides, only the case $C \gg C_0$ was considered. As it will be shown below, for spherical granules it is necessary to use an opposite sign on an inequality. In a limit of small currents average electric field differs from that in absence of tunneling a little. At a homogeneous average field approach of not interacting parallel chains for the three-dimensional granulated medium is represented justified. I.e. the problem is reduced to one chain with an identical average electric field along a chain. In the present work in approach of a homogeneous average field same, as well as in [1] method of Green functions, the problem of discrete electronic transport through a chain of as much as big number of metal granules in area Coulomb tunneling blockade is analytically solved. It is deduced the exponential law of growth of current with electric field in the granulated medium. Examples of application of the theory for an explanation of transport of current in glass like amorphous semiconductors, porous silicon, synthesized by ionic implantation of nitrogen in silicon layers of nitride of silicon are resulted.

Offered work is the corrected and added variant of article published earlier in Russian [8].

## 2. INITIAL ASSUMPTIONS

Let's consider system of spherical bodies - granules metal (or semiconductor) or the same system with impurity atoms instead of granules in the dielectric substance with the wide energy band gap. Full Hamiltonian of systems, as well as in [1,6] we will present in the form of the sum of three terms

$$H = H_{0F} + H_{CF} + H_T , \quad (1)$$

where $H_{0F}$ – Hamiltonian of all set of not co-operating bodies, $H_{CF}$ - correlation contribution the Coulomb interactions, generally depending on electric field in dielectric, $H_T$ - tunnel term. We will assume that the overwhelming contribution to electrostatic energy is introduced by polarization of dielectric in the nearest vicinity of a granule or atom. Energy of polarization is proportional to square of electric field intensity which in turn quadratically falls down with distance r from a granule or atom. It leads to fast decay of electrostatic energy density under the law $\sim 1/r^4$. It gives the ground to neglect electric capacitance between granules and consider capacity of granule as lonely body that means preservation in the equivalent scheme [7] only values $C_0$ and approach soliton of [7] extremely small extent of an order of distance between granules, i.e. transformation soliton to exciton. In same it is possible to be convinced by an estimation of capacitances of two spherical conductors (see task 213 in [9]). At distance between the nearest points of the spheres, equal to their diameter, the cross capacity $C_{12}$ makes only 7 % from value of $C_{11}$ which is almost equal to cross of single-isolated sphere. The chain of granules is represented by the equivalent scheme in fig. 1 where by analogy with [10] the current generator $I_0$=const by the chain ends is included as driving force of charge transfer. Analogue $C_0$ in the scheme [7] is for brevity designated by a symbol $C$.

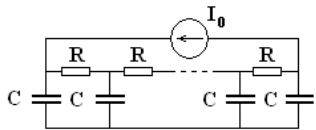

Fig. 1. The equivalent scheme of a chain of granules with tunnel resistance $R$ between them. The role of the common electrode - "ground" carries out an infinitely big and infinitely remote body.

In neglect of cross electrostatic influence of bodies it is possible to write down

$$H_{0F} + H_{CF} = \sum_i (H_0 + H_C)_i , \quad (2)$$

where the index $i$ numbers bodies.

Hamiltonian of each isolated granule with single electron energy of electron bond $E_d$, as well as for atom, it is possible to present in the form of [6]

$$\hat{H}_0 + \hat{H}_C = \sum_{\alpha\sigma} E_d \hat{n}_{\alpha\sigma} + \frac{1}{2}\sum_{\alpha\beta\sigma}(U_{\alpha\beta}-J_{\alpha\beta})\hat{n}_{\alpha\sigma}\hat{n}_{\beta\sigma} + \frac{1}{2}\sum_{\alpha\beta\sigma} U_{\alpha\beta}\hat{n}_{\alpha\sigma}\hat{n}_{\beta,-\sigma}$$
(3)

where the occupation number operator of electrons in shell $\hat{n}_{\alpha\sigma} = a^+_{\alpha\sigma} a_{\alpha\sigma}$, $a^+_{\alpha\sigma}$ and $a_{\alpha\sigma}$ - the the creation and annihilation operators, α, β - sets of quantum numbers (orbital, magnetic), σ - spin number, $U_{\alpha\beta}$ - correlation energy of Coulomb interaction of electrons, $J_{\alpha\beta}$ - energy of exchange interaction.

In neglect of the exchange contribution in comparison with Coulomb interaction and after averaging on orbital and magnetic quantum numbers we will obtain as in [6]

$$H_0 + H_C = E_d \hat{n} + \frac{1}{2}U\hat{n}(\hat{n}-1). \quad (4)$$

The energy operator (4) leads to dependence of average energy on electronic occupation $n$ in form

$$E_n = nE_{d0} + \frac{1}{2}Un(n-1) + const., \quad E_{d0} = E_d + n_{or}U \quad (5)$$

or

$$E_n = \frac{e^2 n^2}{2C} - An + const = \frac{Q^2}{2C} - An + const, \quad (6)$$

where $n_{0r}$ - the initial equilibrium filling defined by level of Fermi, $E_{d0}$, single-electron energy corresponding to this occupation, $C$ - capacitance of a granule or effective electrostatic capacitance of atom, $Q=ne$, $A_{nd}$ – work function. Capacitance $C$ according to (5) is related with correlation energy $U$ by the formula

$$C = (\partial^2 E/\partial Q^2|_{Q=Q_0})^{-1} = e^2/U , \quad (7)$$

work function is

$$A = -E_{d0} + U/2 . \quad (8)$$

For chain $N+2$ of bodies with $N$ intermediate bodies tunnel гамильтониан in (1) in neglect dependence of probability of tunneling on quantum conditions with the account of electronic exchange $T_{i,\,i-1}$ only between the nearest neighbors looks like

$$H_T = \sum_i T_{i,i-1} \sum_{\alpha_{i-1},\alpha_i}(a^+_i a_{i-1} + a^+_{i-1} a_i), \quad (9)$$

where $a^+_i \equiv a^+_{\alpha_i}$, $a_i \equiv a_{\alpha_i}$, i=0, 1, 2 … N+1. Further it is prospective favorable for the brightest manifestation of discrete tunneling a little overlap integrals $T_{i,\,i-1} \ll U$.

We will be limited by low enough temperatures $kT=1/\beta \ll U$. In case of atoms of transition metals $U$ is order or more of 1эВ. Granules are considered small enough that the value exp(U/kT) is much more unity. I.e. considered bodies are the deep many-charge centers in dielectric. We believe, that is available only two the recharge levels - acceptor level $E_{n+1,n}=E_{n+1}-E_n$ and donor level $E_{n,n-1}=E_n-E_{n-1}$ where $n=n_{or}$ and $n_{or}$ corresponds to neutral state of a center. These levels settle down in enough wide band gap of dielectric far from edges of its allowed energy bands not more close than $\Delta E > U \gg kT$. Thus considered centers are amphoteric, can, both to accept, and to deliver electrons. According to (5) $E_{n+1,n} - E_{n,n-1}=U$. The condition $\Delta E > U$ means, that we do not consider hopping conductivity through an electronic exchange center to center and the allowed energy band of dielectric. Let's show, that with such parameters and enough high concentration of the centers in real conditions equilibrium position of Fermi level between $E_{n+1,n}$ and $E_{n,n-1}$ is easily achievable which is necessary for Coulomb blockade of tunneling. While we consider, that external electric field is not present, and neglect by tunnel contribution $H_T$ in (1), that allows applying the equilibrium statistics.

We use the statistical operator

$$\rho = \exp(-(H - \mu \hat{N}), \quad (10)$$

where $\mu$ - Fermi energy, $\hat{N}$ - the operator of full number of electrons. The grand canonical partition function for ensemble of many-charge centers in approach (2) looks like

$$W = Sp\{\rho\} = \prod_i W_i = \prod_i \sum_{n_i} \lambda^{n_i} Z_{n_i}, \quad (11)$$

chemical activity $\lambda=\exp(\beta\mu)$, the internal canonical partition functions defined by initial distribution

$$Z_{n_i} = \sum_{\alpha_{ni}} \exp(-\beta E_{n_i \alpha_{ni}}), \quad (12)$$

subscripts $\alpha_{ni}$ number the basic level and the excited levels of $i$-th center with $n$ electrons.

For the present we will consider all the many-charge centers are identical. According to definition [11] average number of electrons on centers $N=\lambda \partial \ln W/\partial \lambda$ the fraction of full number of centers $N_c$ in the condition with $n$ localized on a center electrons is defined by expression

$$N_c^n = N_c/(1 + \sum_{m \neq n, \alpha_m} g_{nm} \exp\beta(\mu(m-n) - E_{m\alpha_m} + E_{n\alpha_n})), \quad (13)$$

$$g_{nm} = Z_n/Z_m.$$

Tunneling needs to be considered at distances center to center not exceeding value of an order of 10 nanometers, which corresponds high concentration of centers $N_c=(10^{20}-10^{21})$ cm$^{-3}$. The modern technology allows to control easily at such level of an impurity in dielectric so that the concentration of electric active defects is or much less that for the considered centers or by purposeful introduction of impurity of an opposite sign to provide compensation of impurity. Then in dielectric with width band gap $E_g > 2eV$ concentration of intrinsic current carriers is practically equal to zero, under condition of $U\beta >> 1$, according to electroneutrality condition $N_c^+ = N_c^-$ and (13), practically all centers are in the neutral ground state. At small temperatures in the sum of the formula (13) three summand are significant only

$$N_c^n = N_c/(1 + g_{n+1,n} \exp\beta(\mu - E_{n+1,n}) + g_{n,n-1} \exp\beta(E_{n,n-1} - \mu)) \quad (14)$$

with corresponding to neutral state $n=n_{or}$ and actual for discrete tunneling through the neutral centers levels - the first acceptor level $E_{n+1,n}=E_{n+1}-E_n$ and the first donor level $E_{n,n-1}=E_n-E_{n-1}$. At identical effective degeneration factors g of these recharge levels the Fermi level $\mu = -A$ and, according to (8), also settles down precisely in the middle between them.

The carried out consideration of the equilibrium condition is easily generalised for the case of disorder in the sizes of metal or semiconductor granules or for impurity atoms of transition elements. If the centers of gravity between the first acceptor and donor levels in the band gap essentially do not vary with change of diameter of granules for each set of granules of the identical size expression of a kind (14) will be fair and all granules remain mainly in a neutral state.

## 3. THE SOLUTION OF THE PROBLEM FOR THE CHAIN OF GRANULES IN APPROACH OF THE UNIFORM AVERAGE FIELD

In the presence of voltage $V$ on the chain granules appear in electric field $F$. As well as in [1], we will consider the field only through shift of Fermi level $\mu_i$, corresponding to i-th granule (or to i-th atom), i.e. by addition to (2) $H_V = \sum_i eV_i \hat{n}_i$, where potentials $V_i$ are defined in the limit of small currents through chain by geometrical position of granules in the field $F$.

We will consider chain $N+2$ of bodies and demand, as well as in [1], that the statistical operator $\rho$ remained stationary after inclusion of interaction $H_T$ (9). Let $|n_i \alpha_{ni}\rangle \equiv |n_0 \alpha_{n0}, n_1 \alpha_{n1}, ..., n_{N+2} \alpha_{n,N+1}\rangle$ - own functions of Hamiltonian

$$\hat{H} = \hat{H}_{0F} + \hat{H}_{CF} + \hat{H}_V.$$

With these functions $|n_i \alpha_{ni}\rangle$ the big statistical sum (11) is connected

$$Sp\{\rho\}_{\{ni\}} = \sum_{\{\alpha_{ni}\}} \langle n_i \alpha_{ni}|\rho|n_i \alpha_{ni}\rangle = W\{n_i\} \equiv W(n_0, n_1, ..., n_{N+1}),$$
(15)

where - a trace of the operator $\rho$ at the fixed set of occupation numbers $\{ni\} = (n_0, n_1, ... n_{N+1})$, summation is made on set of quantum numbers $\{\alpha_{ni}\} = (\alpha_{n0}, \alpha_{n1}, ..., \alpha_{n,N+1})$.

Further, for the density matrix $\rho'$ with Hamiltonian*

---

* As well as in [1] we do not consider in Hamiltonian in an explicit form the current generator $I_0$. It could be made by analogy to the solving [10] for system of two electrodes by adding as perturbation the contribution from the current generator. Such contribution, again by analogy with [10], will result in occurrence in the kinetic equation for probability density (the equation (16) in this article) of terms $\langle \dot{n}_0 \rangle \partial W'/\partial n_0$ and $\langle \dot{n}_{N+1} \rangle \partial W'/\partial n_{N+1}$ which compensate in average each other according to condition on the preservation of full charge on granules chain.

$$\hat{H}' = \hat{H}_{0F} + \hat{H}_{CF} + \hat{H}_V + \hat{H}_T = \hat{H} + \hat{H}_T$$

with the included contribution of tunneling $H_T$, as well as in [1], we solve Liouville equation $\partial \rho'/\partial t = -i[H',\rho']$, without not conserving number of particles the first order on $H_T$ terms and within square-law on $H_T$ terms. For the considered mode till exit from Coulomb tunneling blockade, when the deviation from equilibrium distribution of electrons is not great, because of $H_T$ smallness it is believed not essential higher orders on this operator and considered in [3,4] splitting or renormalization of energy levels of quantum dots. Applying to the Liouville equation the operation $Sp\{...\}_{\{ni\}}$, we will obtain the system of kinetic equations for distribution functions $W$ of type

$$\partial W'\{n_i\}/\partial t =$$
$$-\left(\sum_{j=1}^{N+1}((P(n_j,n_{j-1})+P(n_{j-1},n_j))+\sum_{j=0}^{N}((P(n_j,n_{j+1})+P(n_{j+1},n_j)))\right)W\{n_i\}+$$
$$+\sum_{j=1}^{N+1}\left(P(n_j-1,n_{j-1}+1)W\{n_i,n_j-1,n_{j-1}+1\}+P(n_j+1,n_{j-1}-1)W\{n_i,n_j+1,n_{j-1}-1\}\right)+$$
$$+\sum_{j=0}^{N}\left(P(n_j-1,n_{j+1}+1)W\{n_i,n_j-1,n_{j+1}+1\}+P(n_j+1,n_{j+1}-1)W\{n_i,n_j+1,n_{j+1}-1\}\right)$$

(16)

where $W'\{n_i\}$ is the perturbed distribution function. The distribution function $W\{n_i, n_j \pm 1, n_{j-1} \mp 1\}$ means $W\{n_i\}$ with occupation number $n_j$ in the set $\{n_i\}$ replaced by corresponding value $n_j \pm 1$. The equations of a kind (16) display change of distribution functions as a result of natural balance of every possible single-electron transition between granules with formation of exciton – a couple with superfluous electron on one granule and hole on the next granule. Frequency of transition $P(n_j,n_{j-1})$, for example, corresponds to a creation of exciton $(n_j-1,n_{j-1}+1)$ as a result of tunnel transition of one electron from a granule $j$ with initial filling $n_j$ on a granule $j-1$ with initial filling $n_{j-1}$, $P(n_{j-1}+1,n_j-1)$ corresponds to annihilation of exciton $(n_{j-1}+1,n_j-1)$ as a result of to tunnel transition of one электрона from a granule $j$-1 with initial filling $n_{j-1}$+1 on a granule $j$ with initial filling $n_j$-1. Contributions with a negative sign in (32) correspond to four variants of an exciton creation.

Frequencies or probabilities of transitions $P(...)$ in (16) are defined by kind expressions

$$P(n_j,n_{j-1}) =$$
$$4T_{j,j-1}^2 \text{Re} \sum_{\alpha_j,\alpha_{j-1}} \int_{-\infty}^{0} d(t-t') <a_{\alpha_{j-1}}(t)a_{\alpha_{j-1}}^+(t')>_{n_j,n_{j-1}} <a_{\alpha_j}^+(t)a_{\alpha_j}(t')>_{n_j,n_{j-1}},$$

$$P(n_j,n_{j+1}) =$$
$$4T_{j,j+1}^2 \text{Re} \sum_{\alpha_j,\alpha_{j+1}} \int_{-\infty}^{0} d(t-t') <a_{\alpha_{j+1}}(t)a_{\alpha_{j+1}}^+(t')>_{n_j,n_{j+1}} <a_{\alpha_j}^+(t)a_{\alpha_j}(t')>_{n_j,n_{j+1}},$$

$$P(n_{j-1}+1,n_j-1) =$$
$$4T_{j,j-1}^2 \text{Re} \sum_{\alpha_j,\alpha_{j-1}} \int_{-\infty}^{0} d(t-t') <a_{\alpha_{j-1}}^+(t)a_{\alpha_{j-1}}(t')>_{n_j-1,n_{j-1}+1} <a_{\alpha_j}(t)a_{\alpha_j}^+(t')>_{n_j-1,n_{j-1}+1},$$

$$P(n_{j+1}+1,n_j-1) =$$
$$4T_{j,j+1}^2 \text{Re} \sum_{\alpha_j,\alpha_{j+1}} \int_{-\infty}^{0} d(t-t') <a_{\alpha_{j+1}}^+(t)a_{\alpha_{j+1}}(t')>_{n_j-1,n_{j+1}+1} <a_{\alpha_j}(t)a_{\alpha_j}^+(t')>_{n_j-1,n_{j+1}+1},$$

(17)

where averages of operators of particles number are defined by the rule

$$<a_{\alpha_j}^+(t)a_{\alpha_j}(t')>_{n_j,n_{j-1}} = Sp\{a_{\alpha_j}^+(t)a_{\alpha_j}(t')\rho\}/Sp\{\rho\},$$
$$<a_{\alpha_{j-1}}^+(t)a_{\alpha_{j-1}}(t')>_{n_j-1,n_{j-1}+1} =$$
$$Sp\{a_{\alpha_j}^+(t)a_{\alpha_j}(t')\rho_{n_j-1,n_{j-1}+1}\}/Sp\{\rho_{n_j-1,n_{j-1}+1}\}.$$

(18)

Other four kinds of probabilities of transitions in (16) are similarly defined. With the same approach of a constancy of density of conditions near Fermi level in each granule, as well as in [1] for one granule, calculation of integrals (17) with transition from summation on momentums $\alpha_i = \bar{k}_i$ on energies gives to integration expressions

$$P(n_j,n_{j-1}) = \lambda_{j,j-1} f(E_{n_j,n_{j-1}} - E_{n_j+1,n_{j-1}} + eV_{j,j-1}),$$
$$P(n_j,n_{j+1}) = \lambda_{j+1,j} f(E_{n_j,n_{j-1}} - E_{n_j+1,n_{j-1}} - eV_{j+1,j}),$$
$$P(n_{j-1},n_j) = \lambda_{j,j-1} f(E_{n_{j-1},n_{j-1}-1} - E_{n_j+1,n_j} - eV_{j,j-1}),$$
$$P(n_{j+1},n_j) = \lambda_{j+1,j} f(E_{n_{j-1},n_{j-1}-1} - E_{n_j+1,n_j} + eV_{j+1,j}),$$
$$P(n_{j-1}+1,n_j-1) = \lambda_{j,j-1} f(E_{n_j+1,n_{j-1}} - E_{n_j,n_{j-1}} - eV_{j,j-1}),$$
$$P(n_{j+1}+1,n_j-1) = \lambda_{j+1,j} f(E_{n_j+1,n_{j-1}} - E_{n_j,n_{j-1}} + eV_{j+1,j}),$$
$$P(n_j+1,n_{j-1}-1) = \lambda_{j,j-1} f(E_{n_j+1,n_j} - E_{n_{j-1},n_{j-1}-1} + eV_{j,j-1}),$$
$$P(n_j+1,n_{j+1}-1) = \lambda_{j+1,j} f(E_{n_j+1,n_j} - E_{n_{j+1},n_{j+1}-1} - eV_{j+1,j}).$$

(19)

where $\lambda_{j,j-1} = 4\pi T_{j,j-1}^2 N_j(0) N_{j-1}(0)$, - density of states at granule Fermi level $j$, $f(x) = x/(1-\exp(-\beta x))$, the potential difference between granules $V_{j,j-1}$ is defined by a difference of Fermi levels of granules $eV_{j,j-1} = \mu_j - \mu_{j-1}$ and differs from zero in the presence of current through granules $j$ and $j$-1. First four expressions (19) define probabilities of exciton formation with superfluous charges $-+, +-$ on the next granules, second four probabilities – annihilation of these excitons.

From the condition $\partial W'(n_i)/\partial t = 0$ we obtain the system of equations of stationarity

$$W\{n_i\} =$$
$$\left(\sum_{j=1}^{N+1}\left(P(n_{j-1}+1,n_j-1)W\{n_i,n_j-1,n_{j-1}+1\}+P(n_j+1,n_{j-1}-1)W\{n_i,n_j+1,n_{j-1}-1\}\right)+\right.$$
$$\left.+\sum_{j=0}^{N}\left(P(n_{j+1}+1,n_j-1)W\{n_i,n_j-1,n_{j+1}+1\}+P(n_j+1,n_{j+1}-1)W\{n_i,n_j+1,n_{j+1}-1\}\right)\right)\times$$
$$\times\left(\sum_{j=1}^{N+1}((P(n_j,n_{j-1})+P(n_{j-1},n_j))+\sum_{j=0}^{N}((P(n_j,n_{j+1})+P(n_{j+1},n_j)))\right)^{-1}.$$

(20)

As the subject of our interest is the regime of small currents before the beginning of exit from a Coulomb tunneling blockade use approach of enough weak electric fields

$$E_{n_{j-1}+1,n_{j-1}} - E_{n_j,n_{j-1}} >> eV_{j,j-1} \qquad (21)$$

and enough low temperatures
$$\exp(E_{n_{j-1}+1,n_{j-1}} - E_{n_j,n_j-1})/kT \gg 1. \quad (22)$$
Let's introduce between the next granules cross correlation energy (CCE)
$$E_{n_j,n_j-1} - E_{n_{j-1}+1,n_{j-1}} = -U_{j,j-1} = E_{n_j,n_j-1} - E_{n_j+1,n_j} \equiv -U_j, \quad (23)$$
For the case when all granules are identical, according to (5) where cross correlation energy (CCE) neighbor granules $U_{j,j-1} = U_j$ and correlation energy does not depend on number of granule $U_j = U$. Under these conditions and identical parameters of tunneling to the left and to the right $\lambda_{j,j-1} = \lambda_{j,j+1} = \lambda$ expression (20) is reduced to the formula

$$W\{n_i\} \approx \frac{1}{2(N+1)} \Bigg( \sum_{j=1}^{N+1} (W\{n_i, n_j-1, n_{j-1}+1\} + W\{n_i, n_j+1, n_{j-1}-1\}) +$$
$$\sum_{j=0}^{N} (W\{n_i, n_j-1, n_{j+1}+1\} + W\{n_i, n_j+1, n_{j+1}-1\}) \Bigg) \exp(U/kT), \quad (24)$$

from which it is visible, that each of no equilibrium values $W\{n_i, n_j \pm 1, n_{j-1} \mp 1\}$ or $W\{n_i, n_j \pm 1, n_{j+1} \mp 1\}$ satisfies to the condition of smallness
$$W\{n_i, n_j \pm 1, n_{j-1} \mp 1\} \approx W\{n_i, n_j \pm 1, n_{j+1} \mp 1\} \approx$$
$$W\{n_i\} \exp(-U/kT) \ll W\{n_i\}. \quad (25)$$

Apparently (24), (25) with a certain average size CCE=$U$ can be carried out and in case of different granules, but at keeping of conditions (37), (38) and $\lambda_{j,j-1} = \lambda_{j,j+1} = \lambda$ with a certain average $\lambda$ for each pair of neighbor granules. Therefore and in connection with monotonous dependence of frequencies of transitions on a voltage difference between granules, at current calculation through granules it is possible to use with a small error the equilibrium statistical sum $W\{n_i\}$ with an initial equilibrium set of numbers of occupation $\{n_{ior}\}$. The average current of granules chain is equal to the current between any pair of neighbor granules. By analogy to the formula (4.2) in [1] current between two neighbor granules is defined by the sum of contributions of tunneling processes along and against the field with every possible weight values $W\{n_i\}$

$$I_{ch} = e \sum_{n_0=n_{or}-1}^{n_{or}+1} \sum_{n_1=n_{or}-1}^{n_{or}+1} \ldots \sum_{n_{N+1}=n_{or}-1}^{n_{or}+1} W\{n_i\}(P(n_j, n_{j-1}) - P(n_{j-1}, n_j)) \approx$$
$$\approx e((P(n_j, n_{j-1}) - P(n_{j-1}, n_j)) = \frac{U}{eR} \exp\left(-\frac{U}{kT}\right) sh\left(\frac{eV_{j,j-1}}{kT}\right) =$$
$$= \frac{e}{R_{j,j-1}C_{j,j-1}} \exp\left(-\frac{e^2}{C_{j,j-1}kT}\right) sh\left(\frac{eV_{j,j-1}}{kT}\right),$$
$$(26)$$

where tunnel resistances are $R_{j,j-1} = 1/\lambda_{j,j-1}e^2$ and, according to (7), electric capacitances are $C_{j,j-1} = 2C_j C_{j-1}/(C_j + C_{j-1})$. In calculations of approximate expressions in (26) conditions (22), (23), (25) are considered and normalization
$$\sum_{n_0=n_{or}-1}^{n_{or}+1} \sum_{n_1=n_{or}-1}^{n_{or}+1} \ldots \sum_{n_{N+1}=n_{or}-1}^{n_{or}+1} W\{n_i\} = 1 \quad (27)$$

is used. The voltage on a chain with $N+2$ granules is equal the sum of voltages between each pair of neighbor granules and volt-ampere characteristic of the chain looks like
$$V = \sum_{j=1}^{N+1} V_{j,j-1} = \sum_{j=1}^{N+1} \frac{kT}{e} arsh\left( I_{ch} \frac{R_{j,j-1}C_{j,j-1}}{e} \exp\left(\frac{e^2}{C_{j,j-1}kT}\right)\right). \quad (28)$$

At small field $eV_{j,j-1} \ll kT$ volt-ampere characteristic of the chain, as well as in [1] for one granule between metal banks, is linear
$$I_{ch} = \frac{eV}{kT} \left[ \sum_{j=1}^{N+1} \left( \frac{R_{j,j-1}C_{j,j-1}}{e} \exp\left(\frac{e^2}{C_{j,j-1}kT}\right)\right) \right]^{-1}. \quad (29)$$

According to (28) volt-ampere characteristic of the chain with dependence of current on voltage at condition $U_{j,j-1} \gg eV_{j,j-1} \gg kT$ (intermediate fields) the exponential field growth of current with voltage takes the place

$$I_{ch} = \left( \prod_{j=1}^{N+1} 2\left( \frac{R_{j,j-1}C_{j,j-1}}{e} \exp\left(\frac{e^2}{C_{j,j-1}kT}\right)\right) \right)^{-\frac{1}{N+1}} \exp\left(\frac{eV}{kT(N+1)}\right) =$$
$$= \left( \prod_{j=1}^{N+1} 2\left( \frac{R_{j,j-1}C_{j,j-1}}{e} \exp\left(\frac{e^2}{C_{j,j-1}kT}\right)\right) \right)^{-\frac{1}{N+1}} \exp\left(\frac{eV}{kT(N_g-1)}\right),$$
$$(30)$$

where $N_g = N+2$ - number of granules in a chain. In a case when all granules are identical
$$I_{ch} = \frac{e}{RC} \exp\left(-\frac{e^2}{CkT}\right) sh\left(\frac{eV}{(N+1)kT}\right) =$$
$$\frac{e}{RC} \exp\left(-\frac{e^2}{CkT}\right) sh\left(\frac{eV}{(N_g-1)kT}\right) \quad (31)$$

At small voltages $eV_{j,j-1} \ll kT$ the volt-ampere characteristic of chain, as well as in [1] for one granule between metal plates, is linear with conductivity
$$G_{ch} = \frac{e^2}{(N_g-1)RCkT} \exp\left(-\frac{e^2}{CkT}\right) =$$
$$= \frac{e^2}{(N_g-1)RCkT} \exp\left(-\frac{U}{kT}\right). \quad (32)$$

Under the condition $U \gg eV_{j,j-1} \gg kT$ (intermediate fields) exponential field growth of current takes place
$$I_{ch} = \frac{e}{2RC} \exp\left(-\frac{e^2}{CkT}\right) \exp\left(\frac{Ve}{(N_g-1)kT}\right) =$$
$$= \frac{e}{2RC} \exp\frac{e}{kT}\left(\frac{V}{N_g-1} - \frac{e}{C}\right). \quad (33)$$

Formulas (31-33) are fair and in the case of not identical granules if conditions (22,23) are carried out for minimum value $U_{j,j-1}$ of corresponding pair of the granules one of which has the maximum size in the chain. Thus certain averaged $R_{j,j-1}$ and $C_j$ carry out role of parametres R and C.

Character of averaging is easy for deducing from these formulas.

The performed consideration corresponds to regime of the current generator when the chain of granules is connected to a source with infinitely big internal resistance. The problem has the simple solving and in the case when ends granules of chain are infinitely great $C_0=C_{N+1}=\infty$, $U_0$, $U_{N+1} \ll kT$, but inequalities (22,23) for all CCE are still carried out, including $U_{1,0}$ and $U_{N+1,N}$ for the first and last pairs of chain. This variant corresponds to the regime of voltage generator. Thus $U_{1,0}=U_{N+1,N}=U/2$ and in formulas (31-33) instead of $N+1$ it is necessary to substitute $N$. The volt-ampere characteristic in small and intermediate fields, conductivity in small fields and exponential current growth in intermediate fields, accordingly, look like

$$I_{ch} = \frac{e}{2RC}\exp\left(-\frac{N_g e^2}{(N_g+1)CkT}\right) sh\left(\frac{eV}{(N_g+1)kT}\right), \quad (34)$$

$$G_{ch} = \frac{e^2}{(N_g+1)RCkT}\exp\left(-\frac{N_g e^2}{(N_g+1)CkT}\right), \quad (35)$$

$$I_{ch} = \frac{e}{2RC}\exp\left(-\frac{N_g e^2}{(N_g+1)CkT}\right)\exp\left(\frac{eV}{(N_g+1)kT}\right). \quad (36)$$

In the presence of the super big granules in a chain the problem, obviously, is reduced to series connection of chains with such granules on the ends.

It is possible to imagine the two-dimensional or three-dimensional granulated medium as system of parallel chains if deviations in sizes and density of spatial distribution of granules are not great and also potential differences between the nearest granules of mentally allocated neighbor chains much less than longitudinal values $V_{j,j-1}$. Then the volt-ampere characteristic and conductivity of system of conducting plates with the granulated medium between them will be described by formulas of a kind (34-36) with the account of cross-section to current of two-dimensional density of granules.

The obtained results can be practically useful to a simple estimation of concentration of granules in a medium. According to the formula (36) for intermediate fields number of granules $N_g$ along the current line can be defined from slope of the plot of dependence of $\ln I(V)$ by expression

$$N_g = \frac{e \lg e}{kT}\left(\frac{d\lg I}{dV}\right)^{-1} - 1 \approx (17 \cdot \left(\frac{300}{T}\right)\left(\frac{d\lg I}{dV}\right)^{-1} - 1), B^{-1}, \quad (37)$$

and the average capacitance $C$ of granules - from temperature shifts at a fixed voltage

$$C = e\left(\left(\frac{V}{N_g+1}\right) - \frac{kT_1 T_2}{e(T_1-T_2)}\ln\frac{I_{T_2}}{I_{T_1}}\right)^{-1}. \quad (38)$$

## 4. COMPARISON WITH EXPERIMENTAL DATA

Let's consider experimental witnessing of manifestation of discrete tunneling in the granulated mediums. According to (33), (36) at average fields in such substances there is exponential dependence of current density on electric field and correspondingly exponential growth of specific conductivity with electric field growth. A characteristic indication of such current transport is the possible great values $N \gg 1$.

Such phenomenon was observed by us at studying of cross-section transport of current in layers of porous silicon (PS) which for the reasons resulted in [12,13] we imagine as system the nanosized granules of silicon interspersed in dielectric dioxide of silicon. As well as in [12] measurements were made on diode structures metal(In)-PS-Si. To exclude the contribution in volt-ampere characteristics of area of a spatial charge in area of a spatial charge adjoining to the PS in Si, the layer of the PS of thickness 2.7 microns was formed, as well as in [12], by anode etching on substrate of monocrystal electronic silicon KES-0.01 strongly doped by antimony. Unlike [12], for stabilization of properties of the PS, silicon substrates with PS layer were exposed to short-term oxidizing annealing at temperature 700ºC during 10 mines on air. Indium contacts in diameter (0.5-1) mm were used. In fig. 2 measured at a room temperature вольтамперные characteristics of two diode structures are shown. Apparently experimental points not bad keep within on straight lines in half-logarithmic scale $\log I \sim V$. Asymmetry of dependences of current on voltage at change of electric polarity, apparently, is connected with induced by a difference of Fermi levels in In and Si electric field in PS and some nonuniformity of porosity of PS layer. The manifestation of the single mechanism of current transport in both directions is the affinity of slopes of plots in the half-logarithmic scale. The slope of half-logarithmic plots about 0.1 V gives, according (37) at room temperature the value granules number along current $N \approx 150$. If to consider, as well as in [11,12], that average distance between silicon granules $\approx 10$ nm we will obtain thickness of area defining dependence $I(V)$ about 1.5 microns, comparable with geometrical cross-section thickness of the porous layer of silicon.

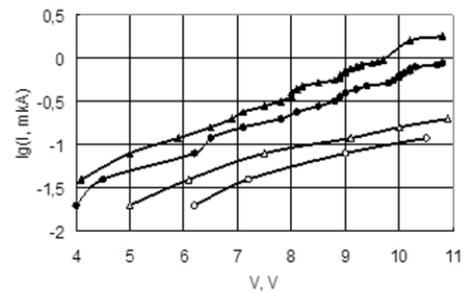

Fig. 2. Volt-ampere characteristics of two diode structures with the layer of porous silicon on n-type silicon single crystal KES 0.01 Ohm·cm. Diode structures differ by contact area. Light points - forward bias, dark - reverse bias branch of volt-ampere characteristics.

The second example of nanogranular mediums is case of layers of silicon nitride formed by ionic implantation of nitrogen in silicon [14]. In such layers with deficiency of nitrogen for synthesis of stoichiometric compound $Si_3N_4$ the formation of silicon nanocrystals, interspersed in a dielectric matrix of nitride of silicon, is possible. The manifestations to it are close to exponential volt-ampere characteristics of diode structures with the layer of silicon nitrated by an ionic-beam method in [14]. Attempts of approximation of dependence $I(V)$ by power function law $I \sim V^n$ according to the theory of injection currents in dielectrics [15] results as well as in case of the PS to too big, not explainable by this theory to power index $n$ from 4 to 6.

Other examples of successful application of the theory for various granulated structures on the basis of silicon are shown in works [16-19].

The next example is voltage dependencies of conductivity which were experimentally observed in amorphous semiconductors. In particular in [20], for thin films of $Te_{48}As_{30}Ge_{12}Si_{10}$ it was obtained at 160K $<T<$500K the exponential voltage dependences of conductivity. Within the physics considered here the amorphous semiconductors similar $Te_{48}As_{30}Ge_{12}Si_{10}$, $Ge_{15}Te_{85}$, $As_2Se_3$ can be considered as nano-granular medium with the narrowband or metal granules in wideband dielectric matrix [8]. This approach allows to explain consistently experimental data and to avoid contradictions the trap model offered by authors [20]. In more details it will be discussed in separate article.

**ACKNOWLEDGEMENTS**

This study was supported by the Russian Foundation for Basic Research (project nos. 08_02_97044ra and 08_02_01222a) and the Federal Agency for Education of the Russian Federation (Rosobrazovanie) (project RNP nos. 2.1.1 4022, 2.1.1/933 and 2.1.1/2833), Federal Purpose Program «Scientific and scientific-pedagogical cadres of innovative Russia», Russian Basic Researches Fund (grants №№ 08-02-97044p, 11-02-00855a).